\documentclass[12pt,a4]{article}

\usepackage{graphicx}
\usepackage{color}

\newcommand{\rf}[1]{(\ref{#1})}
\newcommand{\beq}{\begin{equation}}
\newcommand{\eeq}{\end{equation}}
\newcommand{\bea}{\begin{eqnarray}}
\newcommand{\eea}{\end{eqnarray}}

\newcommand{\e}{{\rm e}}

\newcommand{\Lam}{\Lambda}

\newcommand{\kp}{\kappa}

\newcommand{\define}{\leftdefine}
\newcommand{\leftdefine}{:=}
\newcommand{\rightdefine}{=:}

\newcommand{\Bf}[1]{\mbox{{\boldmath $#1$}}}
\newcommand{\Bfscript}[1]{\mbox{{\scriptsize \boldmath $#1$}}}
\newcommand{\Fm}{F_{\rm m}}
\newcommand{\fm}{f_{\rm m}}

\def\void{}
\def\labelmark{}

\newenvironment{formula}[1]{\def\labelname{#1}
\ifx\void\labelname\def\junk{\begin{displaymath}}
\else\def\junk{\begin{equation}\label{\labelname}}\fi\junk}%
{\ifx\void\labelname\def\junk{\end{displaymath}}
\else\def\junk{\end{equation}}\fi\junk\labelmark\def\labelname{}}

{\ifx\void\labelname\def\junk{\end{array}\end{displaymath}}
\else\def\junk{\end{array}\right.\end{equation}}
\fi\junk\labelmark\def\labelname{}\def\junk{}
}

\newcommand{\beqv}{\begin{formula}{}}

\newcommand{\dd}{{\rm d}}

\newcommand{\tinyspace}{\hspace{0.0556em}}
\newcommand{\dbltinyspace}{\hspace{0.1112em}}
\newcommand{\trpltinyspace}{\hspace{0.1668em}}

\newcommand{\negtinyspace}{\hspace{-0.0556em}}
\newcommand{\negdbltinyspace}{\hspace{-0.1112em}}
\newcommand{\negtrpltinyspace}{\hspace{-0.1668em}}

\vspace{12pt}

\begin{document}

\rightline{\today}

\begin{center}
\vspace{24pt}
{ \Large \bf   The large scale structure of the Universe from a modified Friedmann equation}

\vspace{24pt}

{\sl J.\ Ambj\o rn}$\,^{a,b}$ and {\sl Y.\ Watabiki}$\,^{c}$

\vspace{10pt}

{\small

$^a$~The Niels Bohr Institute, Copenhagen University\\
Blegdamsvej 17, DK-2100 Copenhagen \O , Denmark.\\
email: ambjorn@nbi.dk
\vspace{10pt}

$^b$~Institute for Mathematics, Astrophysics and Particle Physics
(IMAPP)\\ Radbaud University Nijmegen, Heyendaalseweg 135, 6525 AJ, \\
Nijmegen, The Netherlands

\vspace{10pt}

$^c$~Tokyo Institute of Technology,\\ 
Dept. of Physics, High Energy Theory Group,\\ 
2-12-1 Oh-okayama, Meguro-ku, Tokyo 152-8551, Japan\\
{email: watabiki@th.phys.titech.ac.jp}

}

\end{center}

\vspace{24pt}

\begin{center}
{\bf Abstract}
\end{center}

\noindent
We have already shown how a modified Friedmann equation, originating 
from   a model of the Universe built from a certain $W_3$ algebra, is able to explain
the difference between the Hubble constants extracted from CMB data and from local measurements.
In this article we show that the same model also describes aspects of the large scale structure of the Universe well.

\newpage

\section{Introduction}\label{intro}

In previous articles  \cite{aw2,aw3} we have advocated a modified Friedmann equation which is able to resolve the $H_0$ tension and 
where the cosmological constant is assumed to be zero at present time due to Coleman's mechanism \cite{coleman}.
The modified Friedmann equation arose from a ``model of the Universe''  
suggested in \cite{aw1}. One feature of this model is that  the late time 
exponential  expansion of the universe is not caused by the cosmological constant, but by  a new constant 
related to the creation of baby universes.
The possibility of creating such baby universe reflects the fractal structure of spacetime and is not related to 
the vacuum energy dictated by the cosmological constant. 
It is our hope that we will be able to estimate the order of this constant related to the creation of baby universes. 
However, it is not yet possible,
and one should view our modified Friedmann equation as a phenomenological equation with one free parameter, 
$B$, in the same way as the ordinary Friedmann equation has the cosmological constant $\Lambda$ as a 
free parameter. Like $\Lambda$, the parameter $B$
has to be  small in order that the model agrees with observations.
Thus the physics of the modified Friedmann equation as  well as that  of the ordinary Friedmann equation are
independent of $B$ and $\Lambda$ for all 
questions related to observations refering to  times earlier than the time $t_{\rm LS}$ of last scattering.
In this sense our model belongs to a class of models modifying the late time cosmology \cite{latetime,latetime1,lowz}, and in fact our
phenomenological model becomes  similar to models where one puts in by hand a time-dependent dark energy equation of state
for small values of $z$, parametrized in such a way as to ease the $H_0$ or the $S_8$ tension. In these models,
in order to accommodate  larger value of $H_0$ than the value predicted by the Planck data using 
the $\Lam$CDM model \cite{planck}, 
one is led to values of $w(z) < -1$ in some regions of $z$ \cite{latetime1,ps,martin} and since a larger value of $H_0$ 
leads to less growth of density fluctuations for small $z$, 
it has been pointed out that one at the same time might ease the $S_8$ tension \cite{latetime,latetime1}. 

As noticed in \cite{aw3} the situation is the same in our model: our ``effective'' $w(z) < -1$ (and approaches $-1$ for late times),
and the purpose of this article is to show that for precisely the same values of our baby universe parameter $B$ which ease
the  $H_0$ tension, we also agree with a number of other  late time cosmological observations while  assuming the same physics 
as the Planck Collaboration  at and before the time of last scattering (we denote this physics CMB physics).

\section{The modified Friedmann equation and large scale structures}\label{mfequ}

In \cite{aw2} we showed that our modified Friedmann equation  is 
 \beq\label{2.2}
  \left(\frac{\dot{a}}{a}\right)^2 = \frac{\kp \rho }{3} + \frac{B a}{\dot{a}} \frac{(1+3F(x))}{F^2(x)},\quad x := \frac{B a^3}{\dot{a}^3},
  \quad F^2(x)-F^3(x) = x. 
  \eeq
In \rf{2.2} $a(t)$ is the scale factor of the Universe. 
The second term on the rhs of equation \rf{2.2} is a term 
that replaces the term $\Lam/3$ 
which would be present of there was a cosmological constant. 
$x$ has to be less than or equal to 4/27 and 
one has to choose the solution $F(x)$ to the third order equation which is larger than or equal to 2/3.
Despite the missing cosmological constant,
the solution $a(t)$ will grow exponentially for large time as $ a(t) \;\propto\;  \e^{(27 B/4)^{1/3} t}$. However, the replacement 
of $\Lam/3$ with the more complicated term which is a function of $a(t)$ makes it possible to have  physics agreeing with both the 
CMB measurements which refer to  time $t_{\rm LS}$ and the small $z$ measurements which refer to physics at present time. 
In \cite{aw2} we showed that an agreement could be obtained for $H(z)$, 
the Hubble constant as a function of the redshift $z$.
Here we will show that for the values of $B$ determined in \cite{aw2}\footnote{The values of $B$ and the present time $t_0$ 
used in this article are $t_0 = 13.894 \pm 0.057$ Gyr, and $B t_0^3 =0.1492 \pm 0.0056$. They are determined in the same 
way as described in \cite{aw2}, but they are slightly different than the values reported there, 
since we have in this article used the  value  $H_0 =  73.04 \pm 1.04$, which is the latest published value by the SH0ES team
 \cite{latestshoes}}, which led to the resolution of the Hubble constant tension, 
our model, without further modifications, also agrees  with other late time cosmological observations.

\subsection{Baryon Acoustic Oscillations }

The baryon acoustic oscillations (BAO) are 
fluctuations in the baryon density 
caused by the baryon acoustic oscillation in the early universe. 
In BAO an important observable is the inverse of the angular diameter distance 
\begin{equation}\label{angular}
\frac{1}{\theta} \,\define\, \frac{D_{\rm V}(z)}{r_{\rm s}}.
\end{equation}
In this equation $D_{\rm V}(z)$ is a distance measure and $r_{\rm s}$ is the  co-moving sound horizon at $t_{\rm drag}$.
More precisely, 
$D_{\rm V}(z)$ is a kind of the average of 
three directions of distance, i.e.\ 
\begin{equation}
D_{\rm V}(z) \,\define\,
\sqrt[3]{z D_{\rm H}(z) \big( D_{\rm M}(z) \big)^{2}}
\end{equation}
where 
\begin{equation}
z D_{\rm H}(z) \,\define\, \frac{c {\dbltinyspace} z}{H(z)}
\,,
\qquad\quad
D_{\rm M}(z) \,\define\, 
\int_0^{{\tinyspace}z}\! 
  \frac{c {\dbltinyspace} \dd z'}{H(z')}
\,.
\end{equation}
$r_{\rm s}$ is defined by 
\begin{equation}\label{rsDef}
r_{\rm s}
\,=\,
\int_0^{{\dbltinyspace}t_{\rm drag}}\! \dd t {\dbltinyspace}\,
  \frac{a(t_{\rm drag})}{a(t)}
{\dbltinyspace} \,c_{\rm s}(t)
\,,
\end{equation}
where $c_{\rm s}$ is the speed of sound, 
\begin{equation}
c_{\rm s}(t)
\,=\,
\frac{c}{\sqrt{3 \big( 1 + R(t) \big)}}
\,,
\qquad\quad
R(t) \,\define\, 
\frac{3 {\tinyspace} \rho_{{\tinyspace}\rm b}(t)}
     {4 {\tinyspace} \rho_{\gamma}(t)}
\,.
\end{equation}
$\rho_{{\tinyspace}\rm b}(t)$ is the baryon density 
and 
$\rho_{\gamma}(t)$ is the photon density. 
%
%
The baryon drag ends at $t_{\rm drag} < t_{\rm LS}$.
Thus $r_{\rm s}$, as defined by eq.\  \rf{rsDef}, is determined by 
physical phenomena before $t_{\rm LS}$, and according to our assumptions the value of 
$r_{\rm s}$ in our model (which we in the following will label 
with a superscript $(B)$, for the baby universe coupling constant appearing in \rf{2.2}) 
will therefore coincide with the Planck value obtained from the  $\Lambda$CDM model:
\begin{equation}\label{rsValueDrag}
r_{\rm s}^{\rm (B)}  = 
r_{\rm s}^{\rm (SC)}  = 
r_{\rm s}^{\rm (CMB)}
\,=\,
147.05 \pm 0.30
\ [\,{\rm Mpc}\,]
\,.
\end{equation}


\begin{figure}[t]
  \begin{center}
    \includegraphics[width=0.7\linewidth,angle=0]
                    {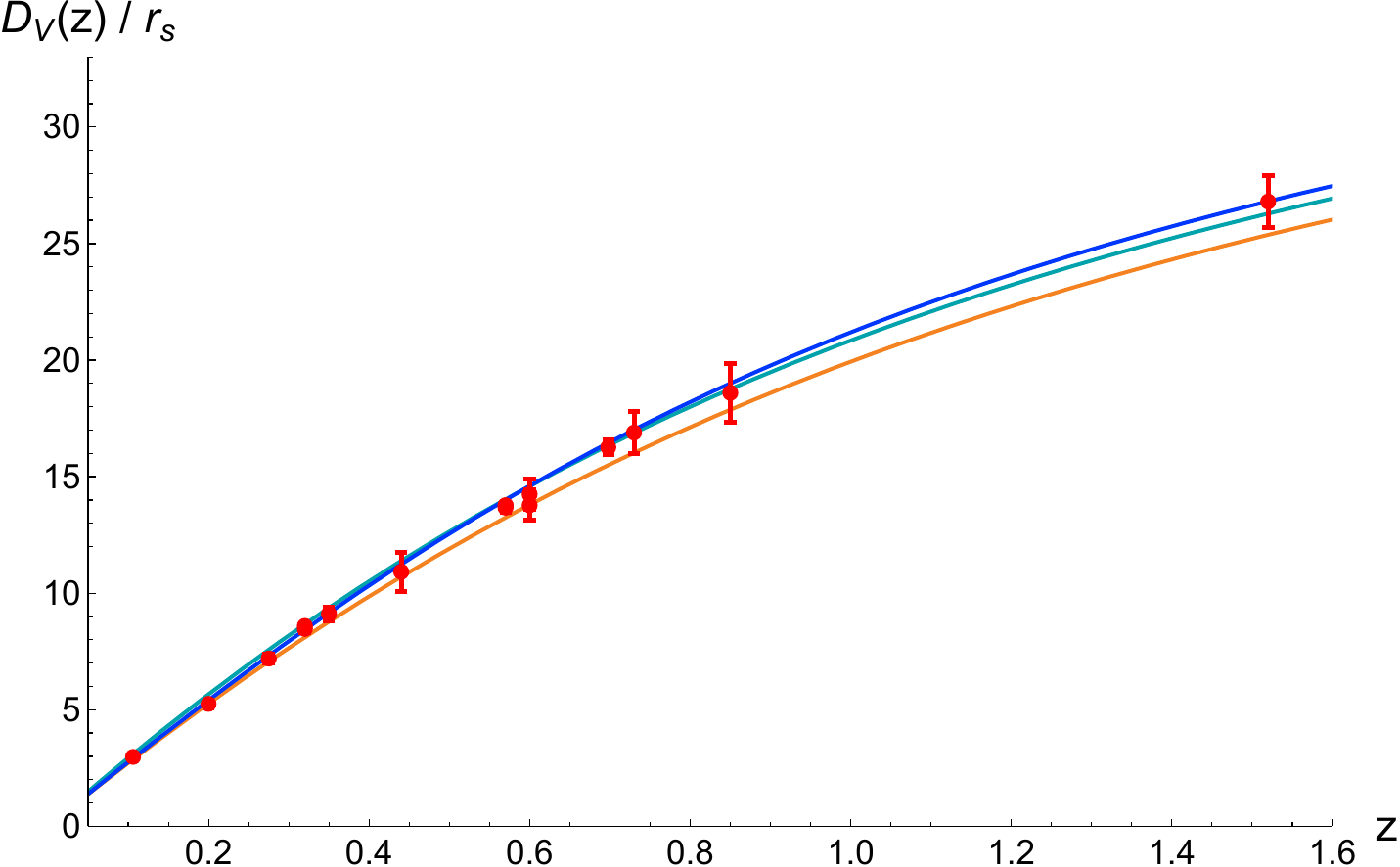}
  \end{center}
\vspace{-10pt}
\caption[Graph_LogHLogH]{\small {$z$ plotted versus {\tinyspace}$D_{\rm V}(z)/r_{\rm s}$ for our model 
(blue), for the Planck data $\Lam$CDM model (green) and the $\Lambda$CDM model with late time $H_0= 73.04$ (orange). 
The data point error bars are from table 1. }
  }
\label{fig:DvRs}
\end{figure}%
Fig.\ \ref{fig:DvRs} shows  $D_{\rm V}(z)/r_{\rm s}$, calculated from our model with $H_0 = 73.04 \pm 1.04$ 
(which, as already mentioned, is the latest value published by the SH0ES team \cite{latestshoes}), for the  $\Lam$CDM model 
with $\Lam$ and $H_0$ given by the Planck collaboration, and for the   $\Lam$CDM model where 
$H_0 = 73.04$ and where the corresponding $\Lam_{\rm SC}$ is
determined from the $\Lam$CDM model by requiring that $z(t_{\rm LS}) = 1089.95$. The reason we consider this third 
model, denoted SC for ``standard candles'' used to measure distances for small $z$, is that it is the $\Lam$CDM model 
where $H(z)$ is calculated using precisely the same assumptions as for our baby universe model.\footnote{%
The values of $\Lambda_{\rm SC}$ and the present time $t_0^{\rm (SC)}$ determined  
in this  way for the SC-$\Lam$CDM model are 
$t_0^{\rm (SC)} = 13.306 \pm 0.051$ Gyr, and 
$\Lam_{\rm SC} t_0^2 =2.167 \pm 0.061$.}
 The data points are  from Table 1 in the Appendix and  lead to reduced $\chi^2$ values\footnote{%
In this article the definition of $( \chi_{\rm red} )^2$ is
$( \chi_{\rm red} )^2 \define
 \sum_{i=1}^N ( (x - x_i) / \sigma_i )^2 / N$, 
where $x_i$ are data with error bars $\sigma_i$. 
}
\begin{equation}
( \chi_{\rm red}^{\rm (B)} )^2
= 1.0
\,,
\qquad
\big( \chi_{\rm red}^{\rm (SC)} \big)^2
= 4.7
\,,
\qquad
\big( \chi_{\rm red}^{\rm (CMB)} \big)^2
= 1.7
\,.
\end{equation}
We note that the  $\Lam$CDM model with $\Lam_{\rm SC}$ and $H_0 = 73.04$ fits the data less well than our model.


\begin{figure}[t]
  \begin{center}
    \includegraphics[width=0.7\linewidth,angle=0]
                    {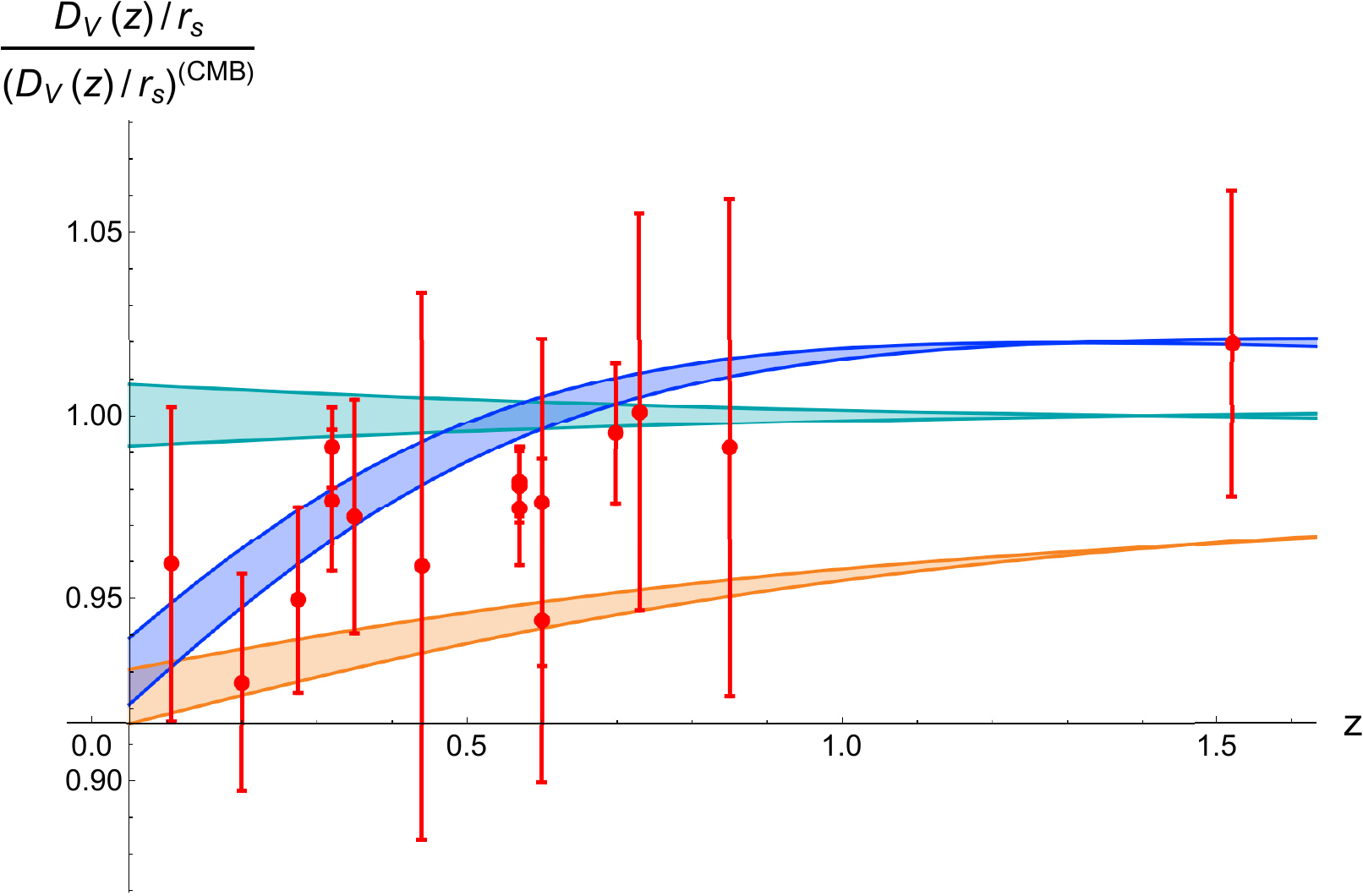}
  \end{center}
\vspace{-10pt}
\caption[Graph_DvRsDiff]{\small 
{$z$ versus $\frac{D_{\rm V}(z)/r_{\rm s}}{D_{\rm V}^{\rm (CMB)}(z)/r_{\rm s}^{\rm (CMB)}}$ for the three models:
our model (blue), Planck data $\Lam$CDM model (green) and late time  $\Lam$CDM model (orange). 
The widths of the curves indicate the $1\sigma$ error bars related to the fits. The error bars of the data are from table 1.}
  }
\label{fig:DvRsDiff}
\end{figure}%

To emphasize the difference for the three models Fig.{\dbltinyspace}\ref{fig:DvRsDiff} shows 
the graphs ${D_{\rm V}(z)}/{r_{\rm s}}$ divided by the central  graph values of 
${D_{\rm V}^{\rm (CMB)}(z)}/{r_{\rm s}^{\rm (CMB)}}$. 
The message to take away is that our model fits the data well.

\subsection{The growth of fluctuations }

The matter density fluctuation 
$\delta_{\rm m}(\Bf{x},t)$ 
is defined by 
\begin{equation}\label{MatterDensityFluctuationDef}
\delta_{\rm m}(\Bf{x},t) \,\define\, 
\frac{\rho_{\rm m}(\Bf{x},t)}{\rho_{\rm m}(t)} - 1
\qquad\hbox{and}\qquad
\rho_{\rm m}(t) \,\define\,
\frac{\int\!\dd^3 x {\dbltinyspace}\rho_{\rm m}(\Bf{x},t)}{\int\!\dd^3 x}
\,.
\end{equation}
In the linear approximation, 
the matter density fluctuation $\delta_{\rm m}(\Bf{x},t)$ 
obeys the differential equation 
\begin{equation}\label{MatterDensityFluctuationDiffeq}
\frac{\dd}{\dd t}\Big(
  \big( a(t) \big)^2 \frac{\dd}{\dd t} \delta_{\rm m}(\Bf{x},t)
\Big)
\,=\,
\frac{\kappa}{2} \big( a(t) \big)^2
  \rho_{\rm m}(t) {\tinyspace} \delta_{\rm m}(\Bf{x},t)
{\dbltinyspace}+{\dbltinyspace}
{\cal O}(B)
\,.
\end{equation}
We here assume that 
$\delta_{\rm m}(\Bf{x},t)$ is decoupled at all times from $t_{\rm LS}$ to the present time $t_0$ as 
\begin{equation}\label{MatterDensityFluctuationDecoupling}
\delta_{\rm m}(\Bf{x},t) \,=\, \Fm(t) \Delta_{\rm m}(\Bf{x})
\end{equation}
where $\Fm(t)$ is the growth factor which represents 
the growth of fluctuation 
and 
$\Delta_{\rm m}(\Bf{x})$ is the fluctuation at a certain time. 
Then we have the following differential equation for $\Fm(t)$:
\begin{equation}\label{MatterDensityFluctuationFtDef}
\frac{\dd}{\dd t}\bigg(
  \big( a(t) \big)^2 \frac{\dd}{\dd t} \Fm(t)
\Big)
=
\frac{\kappa}{2} \big( a(t) \big)^2
\rho_{\rm m}(t) {\tinyspace} \Fm(t)
\,.
\end{equation}
Here we have assumed 
the contribution from the production of baby universes is negligible, 
i.e.\ 
we have deleted ${\cal O}(B)$ in eq.\ \rf{MatterDensityFluctuationDiffeq}. 
The boundary conditions of the growth factor $\Fm(t)$ are 
\begin{equation}\label{MatterDensityFluctuationFtCondition}
\Fm(t_0) \,=\, 1
\qquad\hbox{and}\qquad
\fm(t_{\rm LS})
\,=\,
1
\end{equation}
where 
$\fm(t)$ is the growth rate defined by 
\begin{equation}\label{growthrate}
\fm(t) \,\define\, \frac{\dd \log \Fm(t)}{\dd \log a(t)},
\end{equation}
and where $t_0$ refers to the present time (and $t_{\rm LS}$ as before to the 
time of last scattering).


One now introduces the average of fluctuation 
inside the sphere with radius $R$:
\begin{equation}\label{MatterDensityFluctuationAverage}
\sigma_R(t)
\,\define\,
\sqrt{
\bigg<\!
  \bigg(
    \frac{1}{\frac{4}{3} \pi R^3}
    \!\int_{|\Bfscript{x}| < R}\!\!\!\dd^3 x {\dbltinyspace}
    \delta_{\rm m}(\Bf{x},t)
  \bigg)^{{\negtrpltinyspace}2}
{\trpltinyspace}\bigg>
}
\,\,=\,
\Fm(t)
\sigma_R(t_0),
\end{equation}
where we have used the decoupling property 
\rf{MatterDensityFluctuationDecoupling}. 
Thus we have 
\begin{equation}\label{MatterDensityFluctuationAverage0}
\sigma_R(t_0)
\,=\,
\sqrt{
\bigg<\!
  \bigg(
    \frac{1}{\frac{4}{3} \pi R^3}
    \!\int_{|\Bfscript{x}| < R}\!\!\!\dd^3 x {\dbltinyspace}
    \Delta_{\rm m}(\Bf{x})
  \bigg)^{{\negtrpltinyspace}2}
{\trpltinyspace}\bigg>
}
\,\,\rightdefine\,
\sigma_R\,.
\end{equation}
We have introduced the notation $\sigma_R$ in eq.\ \rf{MatterDensityFluctuationAverage0}
to emphasize that $\sigma_R(t_0)$ is really independent of $t_0$, as is clear from 
the decoupling equation \rf{MatterDensityFluctuationDecoupling}. From this equation 
$\sigma_R$ can only depend on physics before $t_{\rm LS}$ and in our model we have assumed
that the physics before $t_{\rm LS}$ is the CMB physics used by the Planck collaboration.
Therefore, we can write
\begin{equation}\label{sigmaRatCMB}
\sigma_R^{\rm (B)}(t_0)
=
\sigma_R^{\rm (SC)}(t_0)
=
\sigma_R^{\rm (CMB)}(t_0)
=
\sigma_R
\,.
\end{equation}

As above we consider the Planck $\Lambda$CDM model, our modified Friedmann equation model with $H_0=73.04$
and the the $\Lam$CDM model  with $H_0 = 73.04$  (the present time cosmological measurement) 
 and where $z(t_{\rm LS}) = 1089.95$ determines the corresponding $\Lam_{\rm SC}$. 
 
\begin{figure}[t]
  \begin{center}
    \includegraphics[width=0.7\linewidth,angle=0]
                    {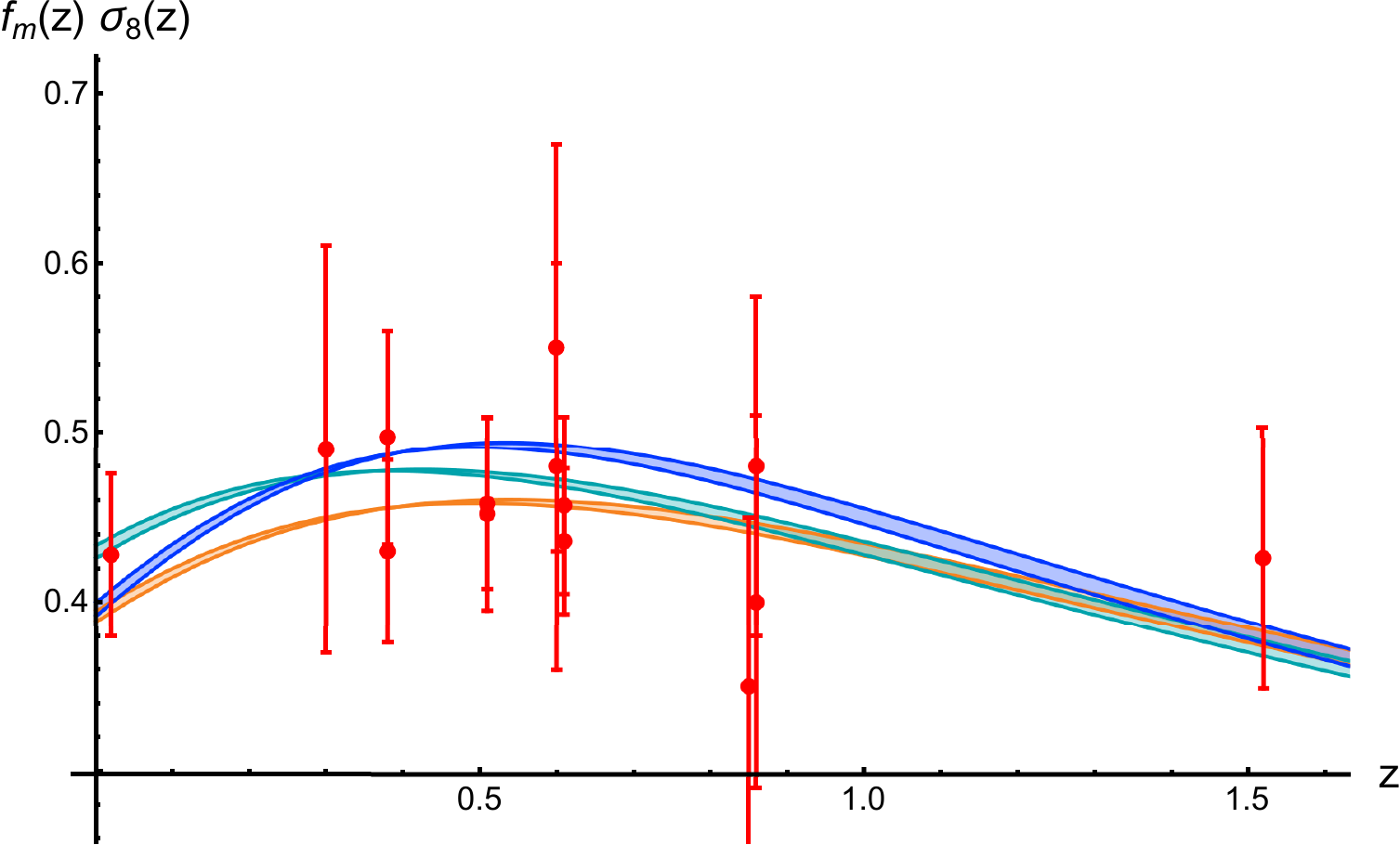}
  \end{center}
\vspace{-10pt}
\caption[Graph_fmsigma8]{\small {$z$ versus $\fm{\negtinyspace}(z) \sigma_8(z)$ for the three models: our model (blue),
 the Planck  $\Lam$CDM 
model (green) and the late time $\Lam$CDM model (orange). 
Data points are from table 2 in the Appendix and the widths of the curves correspond 
to $1\sigma$ coming from the fits.}
  }
\label{fig:fmsigma8}
\end{figure}

 Fig.{\dbltinyspace}\ref{fig:fmsigma8} shows the graphs of 
$\fm{\negtinyspace}(z)\sigma_8(z)$, 
where $\sigma_8(z(t))$ is $\sigma_R(t)$ at 
$R{\negdbltinyspace}={\negdbltinyspace}
8{\tinyspace}{\rm Mpc}/h%
$,\footnote{%
$
8{\dbltinyspace}{\rm Mpc}/h
=
800{\dbltinyspace}{\rm km}{\trpltinyspace}{\rm s}^{-1}
{\negdbltinyspace}/ H_0
$
depends on the present Hubble parameter $H_0$. 
Since $\sigma_R(t)$ is independent of the present time $t_0$, 
the dependence of $t_0$ is artificially introduced in $\sigma_8(t)$. 
So, we here compare with the observed data which use the same $h$ value as the Planck collaboration. 
}
and from \rf{MatterDensityFluctuationAverage} and \rf{sigmaRatCMB} 
it follows that 
\begin{equation}\label{MatterDensityFluctuationAverage8}
\sigma_8(t)
\,=\,
\Fm(t) {\tinyspace}
\sigma_8(t_0) \quad{\rm or} \quad \sigma_8(z(t)) = F_{\rm m}(z(t)) \sigma_8(z_0),
\end{equation}
where $z_0 \define z(t_0) = 0$. Thus
\begin{equation}\label{sigma8atCMB}
\sigma_8^{\rm (B)}(z_0)
=
\sigma_8^{\rm (SC)}(z_0)
=
\sigma_8^{\rm (CMB)}(z_0)
=
\sigma_{R{\dbltinyspace}={\dbltinyspace}8{\rm Mpc}/h}
=
0.8120 \,\pm 0.0073
\end{equation}
for our model. 


The
$( \chi_{\rm red} )^2$ of 
$\fm{\negtinyspace}(t) \sigma_8(t)$
for the three  graph are
\begin{equation}
( \chi_{\rm red}^{(B)} )^2
= 0.49
\,,
\qquad
\big( \chi_{\rm red}^{\rm (SC)} \big)^2
= 0.26
\,,
\qquad
\big( \chi_{\rm red}^{\rm (CMB)} \big)^2
= 0.29
\,.
\end{equation}
The error bars are too large to distinguish between  the models. 

Finally, let us discuss the observable $S_8$, defined as 
\beq\label{S8}
S_8 \define \sigma_8(t_0) ( \Omega_{\rm m}(t_0) / 0.3 )^{0.5}.
\eeq
For this observable there is a tension between the values 
deduced from late time measurements reported in \cite{latetime2} which 
are between 
0.720 and 0.795 
(see Table 3 in the Appendix)
and the value 
\beq\label{s8a}
S_8^{\rm (CMB)} = 0.834 \pm 0.014
\eeq
which is  {\it calculated} from the Planck data using the $\Lam$CDM model. The tension is of the 
order of 3$\sigma$ if one takes into account the error bars. Our value of $S_8$  is 
\beq\label{s8b}
S_8^{\rm (B)} = 
0.769 \pm 0.007
\,,
\eeq
and it agrees well with the  
values deduced from the late time measurements, although our value, like the Planck collaboration value, is 
also {\it calculated}, not measured.



\section{Discussion}

The purpose of this article is to show that our modified Friedmann equation not only resolves the $H_0$ tension 
but also produces a function $H(z)$ which is consistent with the measured 
 large scale structures of the Universe.
We have shown that this is the case for the angular diameter distance defined by eq.\ \rf{angular}, as well as for 
$f_{\rm m}(z) \sigma_8(z)$ defined by 
\rf{growthrate} and \rf{MatterDensityFluctuationAverage8}. 
Finally, our value of 
$S_8$, defined by \rf{S8} and given by \rf{s8b}, agrees with  late time estimates, but has a tension with the value calculated 
from the Planck data by using the $\Lam$CDM model.  Our value, like the Planck value, is also calculated, based 
on physics identical to the CMB physics before $t_{\rm LS}$. 
The reason our $S_8$ value differs from the Planck value of $S_8$ is that our value of $\Omega_{\rm{m},0} = 0.269$
while the Planck value is $\Omega_{\rm{m},0} = 0.317$. The Planck value of $\Omega_{\rm{m},0} H_0^2$ is entirely determined by physics before 
$t_{\rm LS}$ since we have 
\begin{equation}\label{omegaH2}
\frac{\Omega_{\rm m}(t_{\rm LS}) \big( H(t_{\rm LS}) \big)^{{\negtinyspace}2}
    }{\big( 1 + z(t_{\rm LS}) \big)^{{\negtinyspace}3}
    }
\,=\,
\Omega_{\rm m}(t_0) \big( H(t_0) \big)^{{\negtinyspace}2}
\,=\,
\Omega_{{\rm m},0} H_0^2.
\end{equation}
Our model will also satisfy eq.\ \rf{omegaH2} and since our value of $H(t_{\rm LS})$ by assumption agrees
with the Planck value at $z(t_{\rm LS})$ we will have the same value of $\Omega_{{\rm m},0} H_0^2$. However,
by construction we have  a larger 
value for $H_0$ than Planck collaboration since we used the $H_0$ measured at late time to determine 
our value of the constant $B$. Consequently we have a lower value 
of $\Omega_{{\rm m},0}$. 
$\Lam$CDM with $\Lam_{\rm SC}$ has the problem 
in the angular diameter distance.

This is possible because our late time cosmology is governed by our baby universe $B$-term and 
not by the cosmological constant. As already mentioned there are many suggestions of modification of the late time cosmology
which produce more or less the same results as we have reported here for our model. However, many of these models are 
entirely phenomenological in nature, like  using a spline interpolation of the measured function $H(z)$ for small $z$, 
or  modifying the equation of state by hand without too much physical motivation. 
In this sense we feel that our model is special. It has just one parameter, denoted 
$B$ (much like the $\Lam$CDM model), the origin of which goes back to the very early universe where it is related to the 
creation of baby universes. In this sense it relates the very origin of  the Universe to its late destiny (the exponential expansion 
of the Universe at a rate proportional to $\sqrt[3]{B}$), and the presence of the 
parameter $B$  and the modified Friedmann equation is motivated by a physics model of the Universe. Admittedly, 
as already mentioned, we cannot 
presently solve our underlying model in detail and {\it calculate} $B$ and then follow the 
 evolution of the universe from the very beginning of time (which in our model of the Universe is related to the 
 breaking of an underlying $W_3$ symmetry \cite{aw1})   to the present time, but that is 
work in progress....

\vspace{12pt}
\noindent {\bf \large Acknowledgments} 
\vspace{8pt}

YW acknowledges the support from JSPS KAKENHI Grant Number JP18K03612 and  thanks A.\ Hosoya for discussions.

\newpage

\section*{Appendix}


\begin{table}[h]
 \caption{Baryon Acoustic Oscillation}
 \label{table:BaryonAcousticOscillation}
 \centering
  \begin{tabular}{lllll}
   \hline
   $~~z$ & $~~~r_{\rm s}/D_{\rm V}$ & $~~~D_{\rm V}/r_{\rm s}$ &\hspace{-1mm} Survey~~~~~~ & \hspace{-8mm} ArXiv Reference \\
   \hline \hline
    $0.106$ &   $0.336   \pm 0.015$  & & 6dFGS   & 1108.2635   \\
    $0.20$  &   $0.1980  \pm 0.0058$ & & SDSS    & 0705.3323    \\
    $0.275$ &   $0.1390  \pm 0.0037$ & & SDSS    &0907.1660  \\
    $0.35$  &   $0.1097  \pm 0.0036$ & & SDSS    & 1108.2635   \\
    $0.44$  &   $0.0916  \pm 0.0071$ & & WiggleZ & 1108.2635  \\
    $0.60$  &   $0.0702  \pm 0.0032$ & & WiggleZ & 1108.2635  \\
    $0.60$  &   $0.0726  \pm 0.0034$ & & WiggleZ & 1108.2635 \\
    $0.73$  &   $0.0592  \pm 0.0032$~~~ & & WiggleZ & 1108.2635 \\
    $0.57$  & & $13.67   \pm 0.22$     & SDSS    & 1203.6594  \\
    $0.32$  & & $8.4673  \pm 0.1675$   & SDSS    & 1312.4877 \\
    $0.57$  & & $13.7728 \pm 0.1340$   & SDSS    & 1312.4877 \\
    $0.32$  & & $8.59386 \pm 0.09474$  & SDSS    & 1607.03155  \\
    $0.57$  & & $13.7569 \pm 0.14210$~~~  & SDSS    & 1607.03155 \\
    $2.34$  & & $31.2409 \pm 1.08716$  & SDSS    & 1404.1801 \\
    $2.36$  & & $30.3548 \pm 1.08678$  & SDSS    & 1311.1767 \\
    $1.52$  & & $26.8    \pm 1.1$      & SDSS    & 1801.03062   \\
    $0.698$ & & $16.26   \pm 0.31$     & SDSS    & 2007.08993  \\
    $0.85$  & & $18.59   \pm 1.27$     & SDSS    & 2007.09009  \\
   \hline
  \end{tabular}
\end{table}

\vspace{20pt}


\begin{table}[h]
 \caption{Redshift-Space Distortion $f \sigma_8$}
 \label{table:RedshiftSpaceDistortionfsigma8}
 \centering
  \begin{tabular}{llll}
   \hline
   $~~z$ & $~~~f \sigma_8$~~~ & \hspace{-0mm}Survey~~~~ & \hspace{-4mm}ArXiv Reference \\
   \hline \hline
    $0.02$~  & $0.428 \pm 0.048$~~~~~~~~& 6dFGS     & 1611.09862 \\
    $0.30$  & $0.49  \pm 0.12$  & SDSS      & 1310.2820 \\
    $0.38$  & $0.430 \pm 0.054$  & BOSS      & 1607.03148v2 \\
    $0.51$  & $0.452 \pm 0.057$  & BOSS      & 1607.03148v2 \\
    $0.61$  & $0.457 \pm 0.052$  & BOSS      & 1607.03148v2 \\
    $0.38$  & $0.497 \pm 0.063$  & BOSS      & 1607.03155v1 \\
    $0.51$  & $0.458 \pm 0.050$ & BOSS      & 1607.03155v1 \\
    $0.61$  & $0.436 \pm 0.043$ & BOSS      & 1607.03155v1 \\
    $0.60$  & $0.55  \pm 0.12$   & VIPERS    & 1612.05645v3 \\
    $0.86$  & $0.40  \pm 0.11$  & VIPERS    & 1612.05645v3 \\
    $1.52$  & $0.426 \pm 0.077$  & SDSS      & 1801.03062 \\
    $0.85$  & $0.35  \pm 0.10$   &           & 2007.09009 \\
    $0.60$  & $0.48  \pm 0.12$  & VIPERS    & 1612.05647  \\
    $0.86$  & $0.48  \pm 0.10$  & VIPERS    & 1612.05647  \\
   \hline
  \end{tabular}
\end{table}

\vspace{20pt}

\begin{table}[h]
 \caption{$S_8$  
}
 \label{table:LargeScaleStructure}
 \centering
  \begin{tabular}{lll}
   \hline
   $~~~~S_8$ & \hspace{-1mm}Survey~~~~&\hspace{-5mm}  ArXiv Reference \\
   \hline \hline
    $0.784 \pm 0.013$ ~~              &    DES     & 2110.10141 \\
    $0.759 {{+0.024 }\atop{-0.021 }}$& KiDS-1000 & 2007.15633 \\
    $0.759 {{+0.025 }\atop{-0.023 }}$& DES-Y3    & 2105.13543 \\
    $0.795 {{+0.049 }\atop{-0.042 }}$& HSC-BOSS  & 2111.02419 \\
    $0.7781{{+0.0094}\atop{-0.0094}}$& KiDS-DES  & 2105.12108 \\
    $0.766 {{+0.020 }\atop{-0.014 }}$& KiDS-1000 & 2007.15632 \\
    $0.776 {{+0.017 }\atop{-0.017 }}$& DES-Y3    & 2105.13549 \\
    $0.751 {{+0.039 }\atop{-0.039 }}$& BOSS DR12 & 2112.04515 \\
    $0.720 {{+0.042 }\atop{-0.042 }}$& BOSS$+$eBOSS & 2106.12580 \\
    $0.736 {{+0.051 }\atop{-0.051 }}$& BOSS      & 2110.05530 \\
    $0.73  {{+0.03  }\atop{-0.03  }}$& DELS      & 2111.09898 \\
    $0.784 {{+0.015 }\atop{-0.015 }}$& unWISE    & 2105.03421 \\
    $0.78  {{+0.04  }\atop{-0.04  }}$& KiDS-DR3  & 2012.12273 \\
   \hline
  \end{tabular}
\end{table}

\end{document}